\documentclass[aps,pra,twocolumn,superscriptaddress]{revtex4-1}

\usepackage{graphicx}
\usepackage{amsmath}

\begin{document}

\title{High visibility temporal ghost imaging with classical light}

\author{Jianbin Liu}
\affiliation{Electronic Materials Research Laboratory, Key Laboratory of the Ministry of Education \& International Center for Dielectric Research, Xi'an Jiaotong University, Xi'an 710049, China}

\author{Jingjing Wang}
\affiliation{Electronic Materials Research Laboratory, Key Laboratory of the Ministry of Education \& International Center for Dielectric Research, Xi'an Jiaotong University, Xi'an 710049, China}

\author{Hui Chen}
\affiliation{Electronic Materials Research Laboratory, Key Laboratory of the Ministry of Education \& International Center for Dielectric Research, Xi'an Jiaotong University, Xi'an 710049, China}

\author{Huaibin Zheng}
\affiliation{Electronic Materials Research Laboratory, Key Laboratory of the Ministry of Education \& International Center for Dielectric Research, Xi'an Jiaotong University, Xi'an 710049, China}

\author{Yanyan Liu}
\affiliation{Science  and  Technology  on Electro-Optical  Information  Security Control  Laboratory}

\author{Yu Zhou}
\email[]{zhou1@mail.xjtu.edu.cn}
\affiliation{MOE Key Laboratory for Nonequilibrium Synthesis and Modulation of Condensed Matter, Department of Applied Physics, Xi'an Jiaotong University, Xi'an 710049, China}

\author{Fu-li Li}
\affiliation{MOE Key Laboratory for Nonequilibrium Synthesis and Modulation of Condensed Matter, Department of Applied Physics, Xi'an Jiaotong University, Xi'an 710049, China}

\author{Zhuo Xu}
\affiliation{Electronic Materials Research Laboratory, Key Laboratory of the Ministry of Education \& International Center for Dielectric Research, Xi'an Jiaotong University, Xi'an 710049, China}

\date{\today}

\begin{abstract}
High visibility temporal ghost imaging with classical light is possible when superbunching pseudothermal light is employed.  In the numerical simulation, the visibility of temporal ghost imaging with pseudothermal light equaling  ($4.7\pm 0.2$)\% can be increased to ($75\pm 8$)\% in the same scheme with superbunching pseudothermal light. The reasons for the difference in visibility and quality of the retrieved images in different situations are discussed in detail. It is concluded that high visibility and high quality temporal ghost image can be obtained by collecting large enough number of data points. The results are helpful to understand the difference between ghost imaging with classical light and entangled photon pairs. The superbunching pseudothermal light can be employed to improve the image quality in ghost imaging applications.
\end{abstract}

\maketitle

\section{Introduction}\label{introduction}

Ghost imaging is a new imaging technique that is based on the second- or higher-order correlation of light \cite{shih-book}. In a typical ghost imaging scheme, there are two light beams, which are usually emitted by the same light source. One is signal beam and is incident onto an object. The transmitted or reflected light from the object is collected by a bucket detector without position information. The other beam is reference beam, which does not interact with the object and is collected by a scannable detector with position information. The image of the object can not be obtained by either signal from these two detectors. Since the image information is collected by the bucket detector without position resolution. The position resolution of the reference detector is high enough to resolve the object, while the object does not interact with the reference light beam. Surprisingly, the image of the object can be retrieved by correlating the signals from these two detectors. This peculiar property is the reason why the imaging technique is named as ghost imaging \cite{shih-book}.

The developing history of ghost imaging is a typical example of how scientific research advances itself by the efforts of people with different opinions \cite{klyshko,pittman,abouraddy,bennink,gatti-2004,chen-2004,valencia,cai,scarcelli,gatti-comment, scarcelli-reply,shapiro-2008,ragy,boyd-qip,shih-qip,shapiro-qip,boyd-2017}. Ghost imaging was first performed with entangled photon pairs \cite{klyshko,pittman}.  It was then thought that entanglement was necessary for ghost imaging \cite{abouraddy}. However, inspired by the work of Bennink \textit{et al.} \cite{bennink}, it was proved that ghost imaging can also be performed with thermal light \cite{gatti-2004,chen-2004,valencia,cai}. Since then, the discussion about the physics of ghost imaging with thermal light never stops \cite{scarcelli,gatti-comment, scarcelli-reply,shapiro-2008,ragy,boyd-qip,shih-qip,shapiro-qip,boyd-2017}. Scarcelli \textit{et al.} suggested that ghost imaging with thermal light might have to be described quantum mechanically \cite{scarcelli}. Gatti \textit{et al.} disagreed with the conclusion and claimed that classical theory could also be employed to interpret thermal light ghost imaging \cite{gatti-comment, scarcelli-reply}. Shapiro proposed a new type of ghost imaging called computational ghost imaging and pointed out that the nature of ghost imaging is classical \cite{shapiro-2008}. Rogy \textit{et al.} employed quantum discord to study ghost imaging and concluded that there is quantum correlation in thermal light ghost imaging \cite{ragy}. Although there is no final conclusion about the physics of thermal light ghost imaging \cite{shih-book,boyd-qip,shih-qip,shapiro-qip,boyd-2017}, it is well accepted that ghost imaging with thermal light can mimic all the behaviors of ghost imaging with entangled photon pairs except the visibility of the former is lower than the one of the latter \cite{gatti-2004}. The reason is that the degree of second-order coherence, $g^{(2)}(0)$, equals two for thermal light; while it can be much larger than two for entangled photon pairs \cite{loudon-book}. If there is one type of classical light with large value of $g^{(2)}(0)$, high visibility ghost imaging with classical light should also be possible. 

Different methods, such as differential ghost imaging \cite{ferri}, normalized ghost imaging \cite{sun-2012}, correspondence ghost imaging \cite{luo,chen-2013}, high-order ghost imaging \cite{cao-2008,chen-2010,wu-2013}, \textit{etc}, have been employed to improve the visibility of ghost imaging with classical light. However, mathematical algorithms, which require extra calculations, are employed in these methods to improve the visibility  \cite{ferri,sun-2012,luo,chen-2013,cao-2008,chen-2010}. Recently, we have proposed a new type of classical light called superbunching pseudothermal light, in which  $g^{(2)}(0)$ can be much larger than two \cite{liu-arxiv,bai-2017}. With the help of superbunching pseudothermal light, high visibility temporal ghost imaging can be implemented with normal correlation algorithm similar to that in ghost imaging with entangled photon pairs \cite{pittman}. The result contradicts the well accepted conclusion that the visibility is different between ghost imaging with classical and nonclassical light, which should be one important step in the development of ghost imaging and be helpful to understand the physics of ghost imaging. 

Most ghost imaging experiments were in the spatial domain \cite{klyshko,pittman,gatti-2004,chen-2004,valencia,cai,shapiro-2008,ragy,ferri,sun-2012,luo,chen-2013,cao-2008,chen-2010,wu-2013}. Based on the space-time duality \cite{kolner,company,salem}, temporal ghost imaging should also be possible. Shirai \textit{et al.} first proved that temporal ghost imaging with classical pulsed light can be realized by analogy of spatial ghost imaging with thermal light \cite{shirai}.  It was soon realized that temporal ghost imaging can also be performed with entangled photon pairs \cite{cho,dong,denis}, chaotic laser light \cite{chen,ryczkowski}, and chaotic light \cite{devaux-2017,oka,apl-2017}. Recently, computational temporal ghost imaging was also experimentally implemented \cite{devaux}.  There are also two light beams in a typical temporal ghost imaging scheme \cite{shirai,cho,dong,denis,chen,ryczkowski,devaux-2017,oka,apl-2017}. The signal beam is modulated by a temporal object and then is detected by a slow detector, which can not follow the variation of the signal. The role of this detector is the same as that of bucket detector in spatial ghost imaging. The reference beam is directly detected by a fast detector to record the variation of light intensity. The role of the detector is the same as the one of scannable detector in spatial ghost imaging. The image of the temporal object can not be obtained by either signal from these two detectors. However, the image can be retrieved by correlating the signals from these two detectors \cite{ryczkowski}, just like ghost imaging in spatial domain \cite{shih-book}.

The rest of the paper is organized as follows. Theoretical and numerical study of high visibility temporal ghost imaging with superbunching pseudothermal light is in Sec. \ref{theory}. The discussions about why the visibility of temporal ghost imaging can be increased by changing the light source and the difference in the retrieved images in different situations are in Sec. \ref{discussion}. Our conclusions are in Sec. \ref{conclusion}.

\section{Temporal ghost imaging with superbunching pseudothermal light}\label{theory}

\subsection{Theoretical study}

\begin{figure}[htb]
    \centering
    \includegraphics[width=65mm]{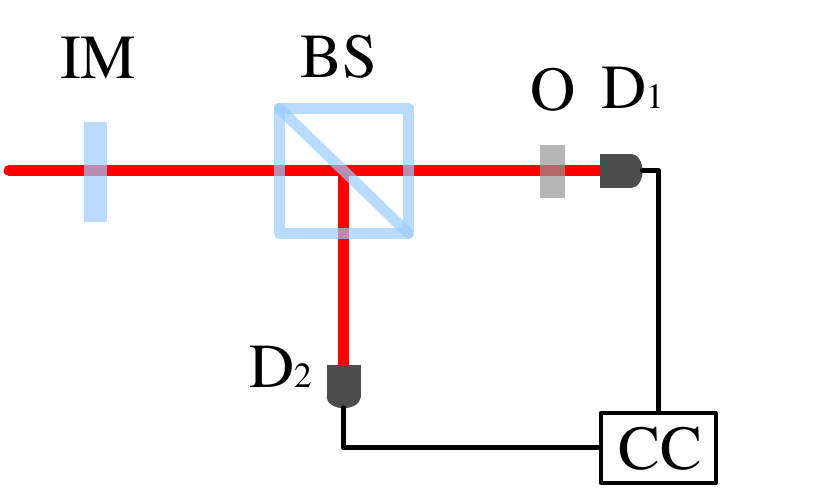}
    \caption{Temporal ghost imaging scheme. IM: intensity modulator. BS: non-polarized $50:50$ beam splitter. O: temporal object. D$_1$: temporal bucket detector, which integrates the intensity within one period. D$_2$: reference detector with high temporal resolution. CC: second-order correlation measurement system. } \label{1}
\end{figure}

The scheme of temporal ghost imaging with superbunching pseudothermal light is shown in Fig. \ref{1}, which is similar to that of spatial ghost imaging with thermal light \cite{scarcelli}. IM is an intensity modulator, which is employed to modulate the intensity of the incident light. It is a combination of rotating ground glasses (RGs), pinholes, lenses, \textit{etc}, in superbunching pseudothermal light source \cite{liu-arxiv,bai-2017}. BS is a non-polarized $50:50$ beam splitter. O is a temporal object, which can be simulated by an electro-optic modulator \cite{ryczkowski}. D$_1$ is a slow detector, which can not resolve the temporal object by itself. D$_2$ is a fast detector, which can follow the modulation of light intensity after IM. CC is a second-order correlation measurement system. The distance between the light source and the temporal object planes is equal to the one between the light source and D$_2$ planes. The difference between spatial and temporal ghost imaging schemes is as follows. In spatial ghost imaging, D$_2$ is scanned transversely to obtain the spatial intensity distribution of the reference light beam. While in temporal ghost imaging, D$_2$ is scanned longitudinally to obtain the temporal intensity fluctuations of the reference light beam. D$_1$ and D$_2$ are in transversely symmetrical positions in temporal ghost imaging so that the same portion of light is received by these two detectors. 

\begin{figure}[htb]
    \centering
    \includegraphics[width=80mm]{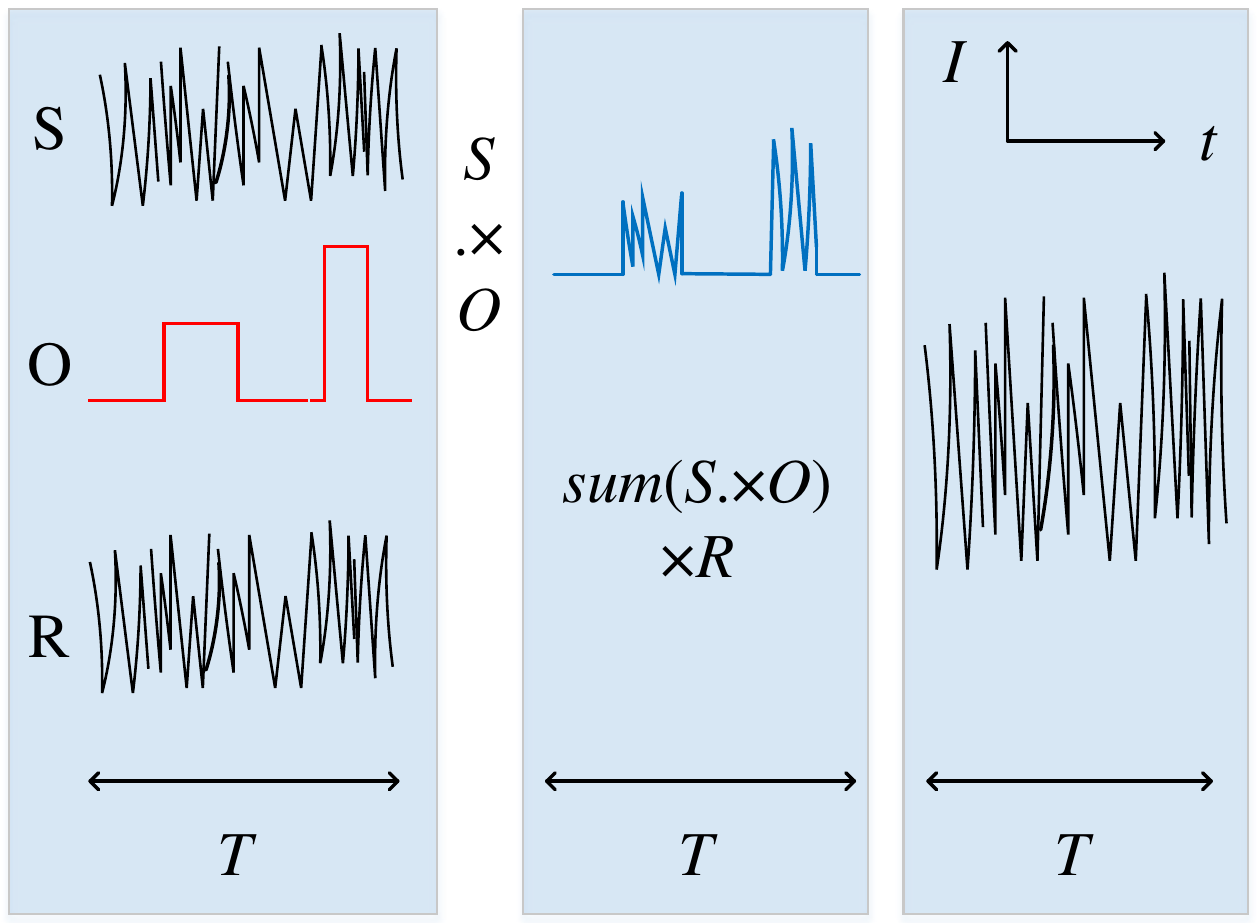}
    \caption{Process of temporal ghost imaging within one period. The horizontal and vertical axes are time ($t$) and intensity of light ($I$), respectively. S: signal beam. O: temporal object. R: reference beam. $T$: period length of temporal object. $S. \times O$: the dot product of signal light beam and temporal object, which corresponds to modulating the temporal object with the signal light beam. $sum(S. \times O)$ is the out of the bucket detector within one period, which is a non-negative real number. $sum(S. \times O) \times R$ represents the intensity of the reference beam multiplied by the output of the bucket detector.} \label{2}
\end{figure}

The process of temporal ghost imaging within one period is described in Fig. \ref{2}. S and R represent signal and reference light beams, respectively. The intensity fluctuations of these two light beams are identical. O is a temporal object, which is a periodic double-slit in our scheme. The period of the temporal object is $T$, which is shown in the left panel of Fig. \ref{2}.  $S. \times O$ means modulating the temporal object with the signal light beam. In the modulating process, the variation of the signal beam should be faster than the temporal object. Otherwise, the information about the object may be lost \cite{shirai}. The modulating process is shown in the middle panel of Fig. \ref{2}.  $Sum(S. \times O)$ is the output of the bucket detector, D$_1$, which integrates the modulated signals within a period. The second-order correlation between signal and reference light beams is implemented by $sum(S. \times O) \times R$.  The image of the temporal object can be retrieved by correlating the outputs of D$_1$ and D$_2$ and repeating the process in Fig. \ref{2} many times \cite{shirai,cho,dong,denis,chen,ryczkowski,devaux-2017,oka} .

The temporal ghost imaging with classical light in Fig. \ref{2} can be mathematically expressed as 
\begin{eqnarray}\label{g2-1}
G^{(2)}(t_2)=  \langle I(t_2) \int_{0}^{T}I(t_1)O(t_1)dt_1\rangle,
\end{eqnarray}
where $I(t_2)$ is the intensity of the reference light beam recorded by D$_2$, $I(t_1)$ is the intensity of the signal light beam,  $O(t_1)$ is the temporal object, and $\int_{0}^{T}I(t_1)O(t_1)dt_1$ is the output of D$_1$ within one period. $\langle...\rangle$ is ensemble average, which is equivalent to a long time average for an ergodic and stationary process \cite{mandel-book}. The temporal object remains the same in different periods. Hence Eq. (\ref{g2-1}) can be re-arranged as
\begin{eqnarray}\label{g2-2}
G^{(2)}(t_2)=  \int_{0}^{T} \langle I(t_2) I(t_1) \rangle O(t_1)dt_1,
\end{eqnarray}
where $\langle I(t_2) I(t_1) \rangle$ is the second-order correlation function \cite{mandel-book}. For a stationary process, the second-order correlation function is only dependent on the time difference between these two intensities \cite{mandel-book}. Equation (\ref{g2-2}) can be simplified as 
\begin{eqnarray}\label{g2-3}
G^{(2)}(t_2)=  \int_{0}^{T} G^{(2)}(t_2-t_1) O(t_1)dt_1,
\end{eqnarray}
where $G^{(2)}(t_2-t_1)$ is the second-order correlation function \cite{glauber,glauber1}. Equation (\ref{g2-3}) is similar to that of lensless spatial ghost imaging with thermal light \cite{scarcelli,shih-book} and is consistent with the results of temporal ghost imaging with classical light \cite{shirai,chen}. 

If the superbunching pseudothermal light proposed in \cite{liu-arxiv,bai-2017} is employed in temporal ghost imaging, Eq. (\ref{g2-3}) can be expressed as
\begin{eqnarray}\label{g2-4}
G^{(2)}(t_2)=  \int_{0}^{T}  \prod_{j=1}^{N} [1+ \text{sinc}^2 \frac{\Delta\omega_j(t_{2}-t_{1})}{2}] O(t_1)dt_1,
\end{eqnarray}
where $N$ is the number of RGs in superbunching pseudothermal light source, $\Delta \omega_j$ is the frequency bandwidth of pseudothermal light scattered by the $j$th RG, and $\text{sinc}(x)$ equals $\sin(x)/x$ \cite{bai-2017}. Equation (\ref{g2-4}) becomes temporal ghost imaging with thermal light when $N$ equals one. The maximal visibility for both temporal and spatial  ghost imaging with thermal light is 33\% \cite{gatti-2004,shih-book}, which is determined by that the degree of second-order coherence of thermal light equals two. The visibility of temporal ghost imaging will increase as $N$ increases.

\subsection{Numerical simulations} 

The intensity of superbunching pseudothermal light follows certain probability distribution, which is dependent on the number of RGs in the superbunching pseudothermal light source \cite{liu-arxiv}. If single-mode continuous-wave laser light is employed as the input of superbunching pseudothermal light source, the intensity of the incident light before the first RG is constant. The intensity, $I$, after the first RG follows negative exponential distribution \cite{goodman-book}
\begin{eqnarray}\label{g2c1}
P_{I|x}(I|x)=\frac{1}{x}\text{exp}(-\frac{I}{x}),
\end{eqnarray}
where $x$  is a constant and is proportional to the intensity of the incident light. If the generated pseudothermal light within one coherence area is filtered by a pinhole and set as the incident light of the second RG, the intensity after the second RG follows a conditional negative exponential distribution, which is conditioned on the intensity of incident light before the second RG. The intensity distribution of superbunching pseudothermal light after two RGs is \cite{goodman-book}
\begin{eqnarray}\label{g2c2}
P_{I}(I)&=&\int_{0}^{\infty}P_{I|x}(I|x)P_x(x)dx \nonumber \\ 
&=&\int_{0}^{\infty}\frac{1}{x}\text{exp}(-\frac{I}{x})\cdot \frac{1}{\langle I \rangle }\text{exp}(-\frac{x}{\langle I \rangle })dx,
\end{eqnarray}
where $P_x(x)$ is the intensity distribution after the first RG and $\langle I \rangle $ is the average intensity of the scattered light after the first RG. 

A third RG can be added after the second RG and the intensity after the third RG is a conditional negative exponential distribution based on the intensity distribution given by Eq. (\ref{g2c2}). Treating  $P_I(I)$ in Eq. (\ref{g2c2}) as $P_x(x)$ and repeating the process in Eq. (\ref{g2c2}) will give the probability distribution of the intensity after the third RG. The process can be repeated for $N$ RGs. The intensity distribution for any number of RGs in the superbunching pseudothermal light source can be obtained in the same way \cite{liu-arxiv}. 

The numerical simulation process of temporal ghost imaging is similar to that in Fig. \ref{2}. 100 random numbers following certain distribution are generated by the method above to simulate intensity fluctuations of superbunching pseudothermal light. Two copies of the random numbers are employed to represent the intensities of the signal and reference light beams, respectively. The intensity of the signal beam is modulated by the temporal object. The output of the bucket detector, D$_2$, is obtained by summing the modulated signals within one period. Then the intensity of the reference light beam is multiplied by the output of D$_2$. The image of the temporal object is retrieved by repeating the process $10^5$ times. 

\begin{figure}[htb]
    \centering
    \includegraphics[width=85mm]{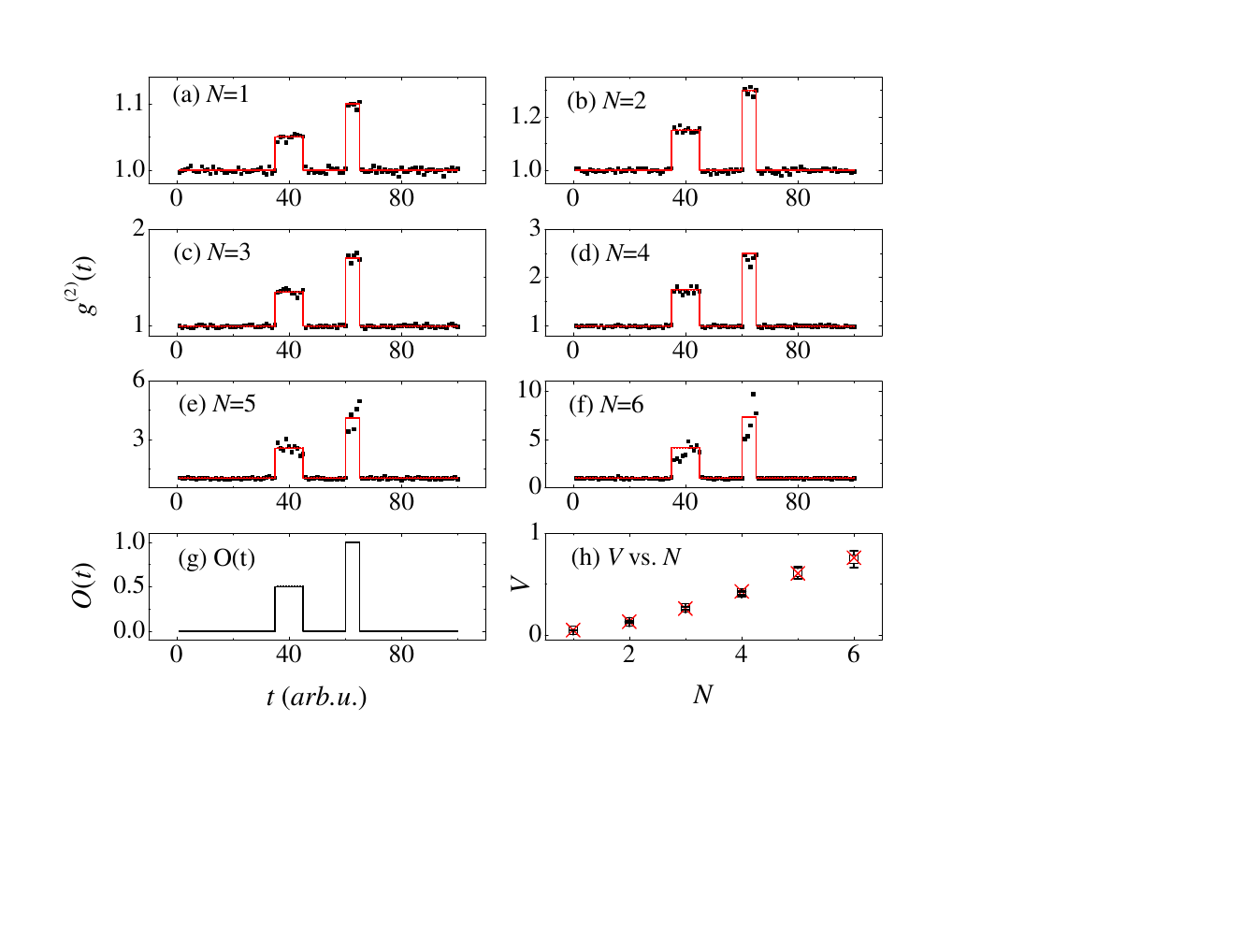}
    \caption{Simulated temporal ghost imaging with superbunching pseudothermal light. $g^{(2)}(t)$ is the normalized second-order correlation function, which is normalized according to the background. $O(t)$ is the temporal object. $t$ is time and $arb.u.$ is short for arbitrary unit of time. The black squares are simulated results and the red lines are theoretical results based on Eq. (\ref{g2-4}). $V$ is visibility and $N$ is the number of RGs in superbunching pseudothermal light source. (a)-(f) are the retrieved images of the temporal object in (g) for different number of RGs in superbunching pseudothermal light source. (h) is the visibility of temporal ghost images in (a)-(f). The hollow black squares are simulated results and red product signs are theoretical values.} \label{3}
\end{figure}

Figure \ref{3} shows the numerical results of temporal ghost imaging with superbunching pseudothermal light. The temporal object is shown in Fig. \ref{3}(g). $O(t)$ represents the amplitude of the temporal object. $t$ is time and $arb. u.$ is short for arbitrary unit of time. The width and height of the first temporal slit are 10 and 0.5, respectively. The width and height of the second temporal slit are 5 and 1, respectively. All the other points are zero. Figures \ref{3}(a)-\ref{3}(f) are the simulated temporal ghost images of temporal object with superbunching pseudothermal light when $N$ equals 1, 2, 3, 4, 5, and 6, respectively. The black squares are numerical results. The red lines in Figs. \ref{3}(a)-\ref{3}(f) are theoretical results by employing Eq. (\ref{g2-4}) when $N$ equals 1, 2, 3, 4, 5, and 6, respectively. $g^{(2)}(t)$ is the normalized second-order correlation function. The meaning of the horizontal axis in Figs. \ref{3}(a)-\ref{3}(f) is the same as the one in Fig. \ref{3}(g). When $N$ equals 1, the scattered light is pseudothermal light \cite{martienssen}. The visibility of temporal ghost image with pseudothermal light in Fig. \ref{3}(a) is ($4.7\pm 0.2$)\%, which is calculated by employing
\begin{eqnarray}\label{vis-d}
V = \frac{g^{(2)}_{max}-g^{(2)}_{min}}{g^{(2)}_{max}+g^{(2)}_{min}}.
\end{eqnarray}
$g^{(2)}_{max}$ is the maximal value of the normalized second-order correlation function and is calculated by averaging the data points at the peak of the second temporal slit in the retrieved image. $g^{(2)}_{min}$ is the minimal value and is calculated by averaging the data points in the background. The visibility of temporal ghost image increases as $N$ increases, which can be seen from the results in Fig. \ref{3}(h).  $V$ is the visibility of temporal ghost image and $N$ is the number of RGs in superbunching pseudothermal light source.  The hollow black squares are numerical results. The red product signs are theoretical results by employing Eq. (\ref{g2-4}). The visibility of temporal ghost imaging with superbunching pseudothermal light can be much higher than the maximal visibility of ghost imaging with thermal light, 33\% \cite{shih-book}. For instance, the visibility in Fig. \ref{3}(f) is ($75\pm 8$)\%. The numerical results are consistent with the theoretical results in Fig. \ref{3}. It is predicted from Eq. (\ref{g2-4}) that the visibility of temporal ghost imaging with superbunching pseudothermal light can approach 1 when $N$ further increases.

\section{Discussions}\label{discussion}

In last section, we have theoretically and numerically proved that high visibility temporal ghost imaging with superbunching pseudothermal light is possible. Comparing the results in Figs. \ref{3}(a)-\ref{3}(f), it is easy to see that the retrieved images in Figs. \ref{3}(a)-\ref{3}(d) are consistent with the temporal object while the retrieved images in Figs. \ref{3}(e)-\ref{3}(f) are somewhat distorted. In this section, we will discuss why there is difference for the retrieved images in different situations and how to eliminate the image distortion. 

\begin{figure}[htb]
    \centering
    \includegraphics[width=75mm]{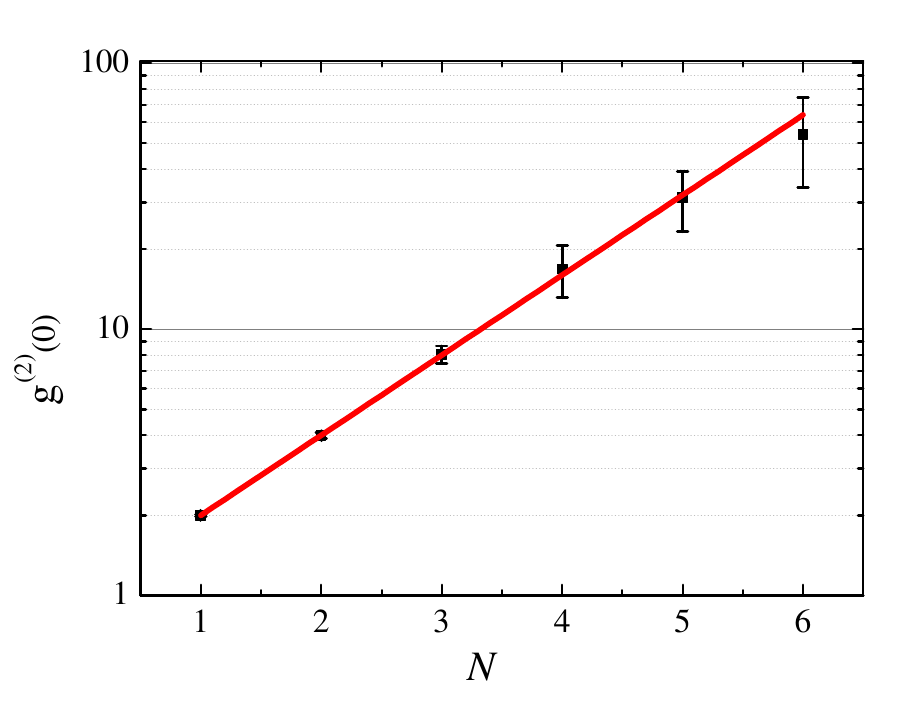}
    \caption{The degree of second-order coherence vs. the number of RGs in superbunching pseudothermal light source. $g^{(2)}(0)$ is the degree of second-order coherence and $N$ is the number of RGs. The black squares are the simulated results and the red line is the theoretical value. The scale of y axis is logarithmic instead of linear.} \label{4}
\end{figure}

Figure \ref{4} shows that the degree of second-order coherence increases with the number of RGs in the superbunching pseudothermal light source. $g^{(2)}(0)$ is the degree of second-order coherence and $N$ is the number of RGs. The black squares are simulated results, which are calculated by generating $2\times 10^4$ random numbers following certain statistics.  The error bar is the standard deviation (SD) and is calculated with 50 independent runs for each data point. The red line is the theoretical value of $2^N$. The results in Fig. \ref{4} can be employed to explain why the quality of the retrieved images are different in Figs. \ref{3}(a)-\ref{3}(f). The lengths of the error bars for $N$ equaling 1 and 2 are less than the size of the black square, which indicates that the fluctuations of the calculated second-order correlation function are small in different runs. As the number of RGs in superbunching pseudothermal light source increases, the SD of the calculated degree of second-order coherence increases. Hence the fluctuation of the calculated second-order coherence function will increase as $N$ gets bigger. This is the reason why the retrieved images are consistent with the temporal object in Figs. \ref{3}(a)-\ref{3}(d) and are distorted in Figs. \ref{3}(e)-\ref{3}(f). 

\begin{figure}[htb]
    \centering
    \includegraphics[width=85mm]{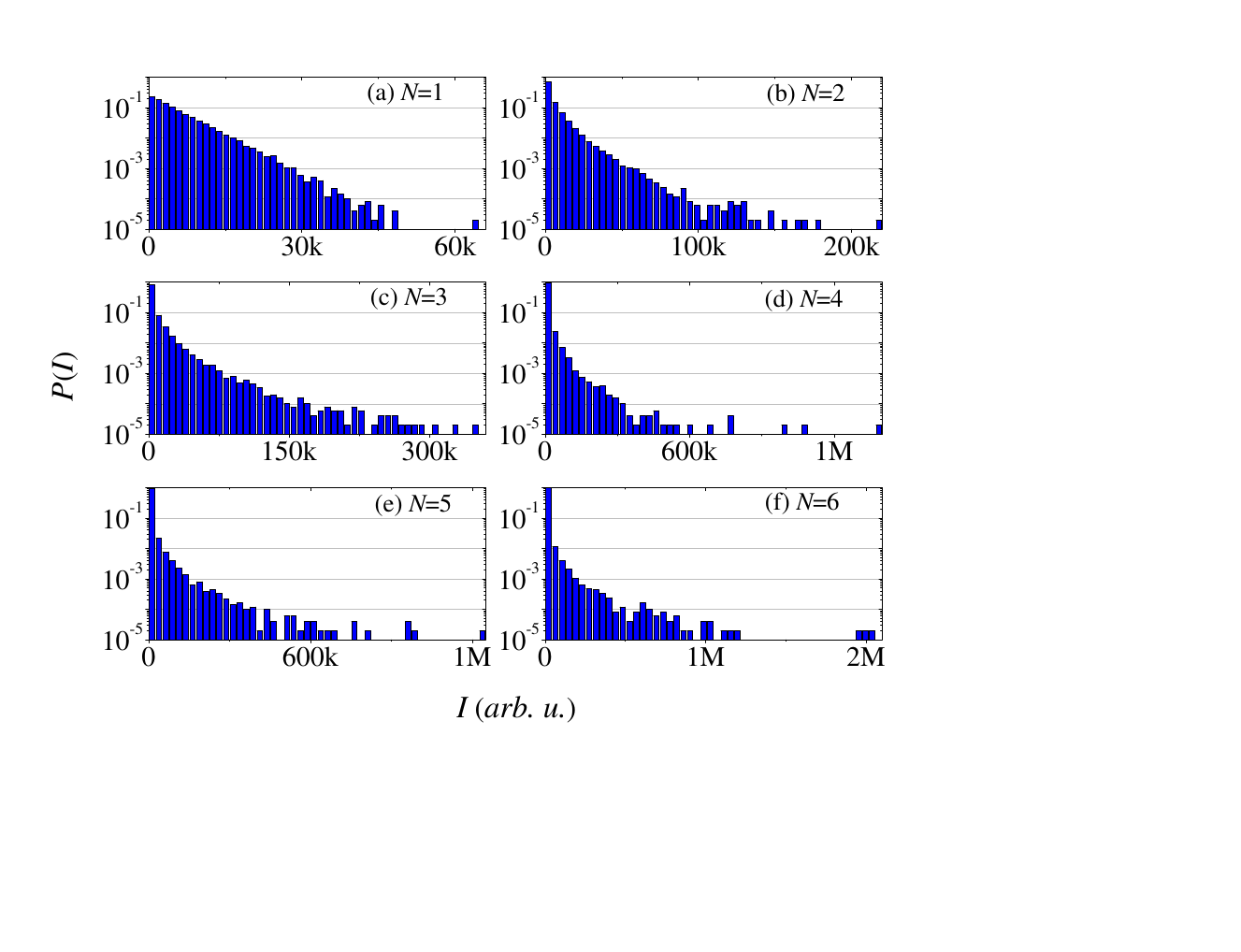}
    \caption{Histograms of numerically generated intensities for different number of RGs in superbunching pseudothermal light source. $5\times10^4$ generated intensities are employed to draw the histogram in each sub-figure. $P(I)$ is the probability for the intensity being $I$ and $I$ is the intensity of light. The average of the generated intensities is 5k and $arb. u.$ is short for arbitrary unit of light intensity.} \label{5}
\end{figure}

The reason why the SD of the calculated degree of second-order coherence increases with $N$ can be understood as follows. Figure \ref{5} shows the histograms of the generated random intensities for different number of RGs in superbunching pseudothermal light source. All the results in Fig. \ref{5} are drawn based on $5\times 10^4$ generated random intensities following certain statistics. When $N$ equals 1, the intensity follows negative exponential distribution, which is shown in Fig. \ref{5}(a). The probability, $P(I)$, decreases as the intensity, $I$, increases. It is possible to obtain enough number of intensities to calculate the degree of second-order coherence with  $2\times 10^4$ random numbers following the same distribution. Hence the calculated value with a limited number of random intensities will be close to the theoretical one when $N$ equals one, as is shown in Fig. \ref{4}. When the number of RGs increases, $P(I)$ mainly concentrates on the far left column in the histogram. For instance, the probability for $I$ being in the far left column equals 0.98 in Fig. \ref{5}(f). The calculated value with the same number of random intensities will fluctuate a lot in different runs. Hence as the number of RGs increases, the SD of the calculated degree of second-order coherence increases, too. The results in Fig. \ref{5} are consistent with the ones in Fig. \ref{4}.

\begin{figure}[htb]
    \centering
    \includegraphics[width=85mm]{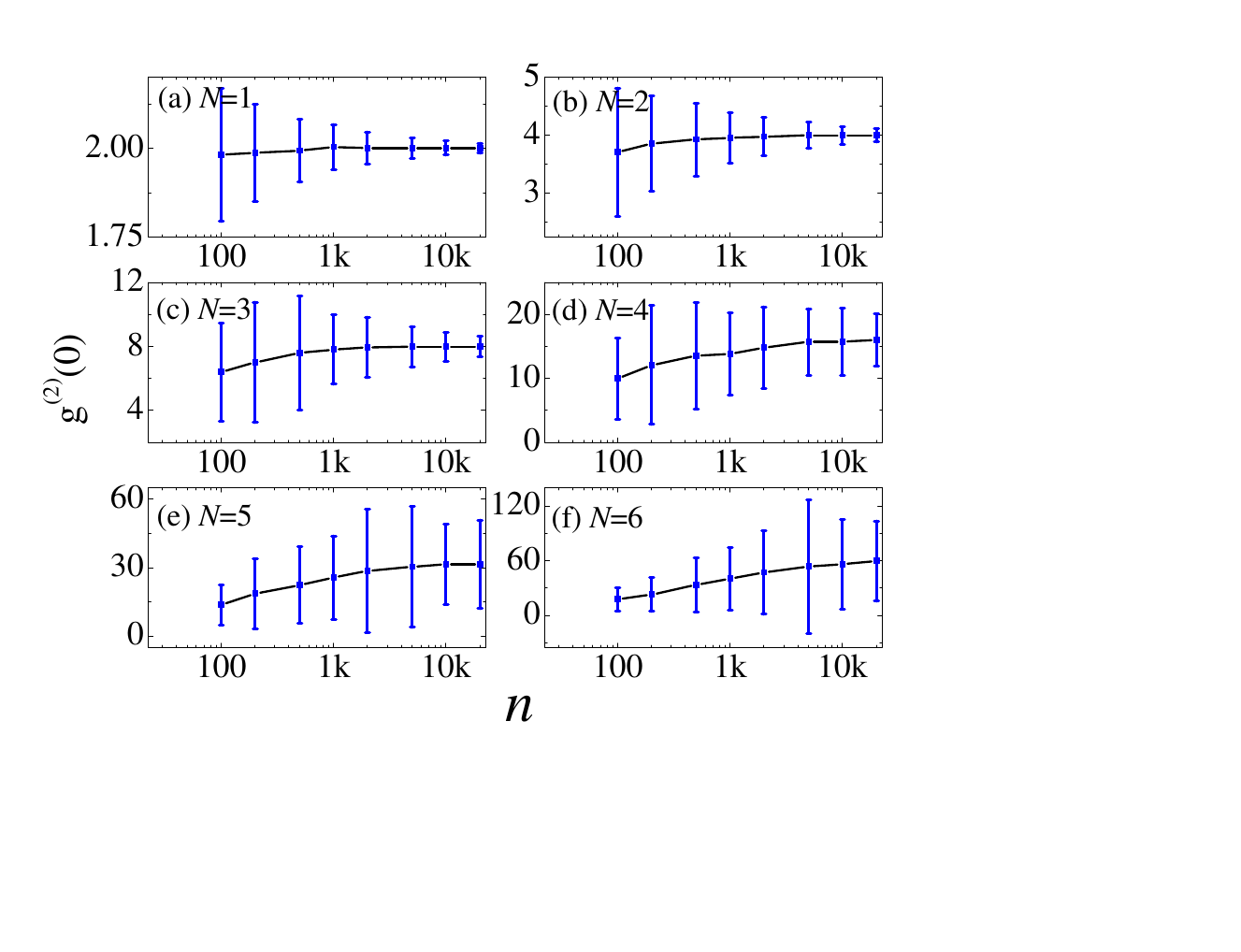}
    \caption{Standard deviation vs. the number of random intensities in a single run. $g^{(2)}(0)$ is the degree of second-order coherence and $n$ is the number of random intensities in a single run. 1000 groups of data are generated to calculate the SD for each data point. The black lines connecting the data points are served as the guidance for the eye. (a)-(f) are the results for 1, 2, 3, 4, 5, and 6 RGs in the superbunching pseudothermal light source, respectively.} \label{6}
\end{figure}

The SD of the calculated degree of second-order coherence can be decreased by increasing the number of the generated random intensities. Figure \ref{6} shows how the SD changes with the number of random intensities for different number of RGs in superbunching pseudothermal light source. The degrees of second-order coherence in Fig. \ref{6} are calculated for 100, 200, 500, 1000, 2000, 5000, 10000, and 20000 random intensities, respectively. The SD is calculated with 1000 different runs for each data point. $N$ is the number of RGs in superbunching pseudothermal light source and $n$ is the number of random intensities in a single run.  The SD in Figs. \ref{6}(a)-\ref{6}(b) decreases as $n$, the number of random intensities, increases. When $N$, the number of RGs in superbunching pseudothermal light source, increases, the calculated SD fluctuates when $n$ increases. For instance, the calculated SD in Figs. \ref{6}(c)-\ref{6}(d) first increases when $n$ increases from 100 to 200 and decreases since then. While the SD in Figs. \ref{6}(e)-\ref{6}(f) increases when $n$ increases from 100 to 5000.

The results in Figs. \ref{4}-\ref{6} can be understood based on the definition of the degree of second-order coherence \cite{mandel-book},
\begin{eqnarray}\label{g2-6}
g^{(2)}0)= \frac{\langle I^2 \rangle}{\langle I \rangle^2},
\end{eqnarray}
where the intensities detected by two detectors are assumed to be identical and $\langle ... \rangle$ is ensemble average. Ensemble average means taking all the possible realizations into account \cite{mandel-book}. When there are enough number of random intensities to calculate the degree of second-order coherence, the calculated result will be close to the theoretical value. When there are not enough number of random intensities, the calculated result is a partial ensemble average and the calculated result may be very different from the theoretical one. Different distributions require different number of random intensities to obtain the results close to the theoretical ones. For instance, $20000 \times 1000$ intensities are enough calculate $g^{(2)}(0)$ for one RG in superbunching pseudothermal light source, which can be seen from the last data point in Fig. \ref{6}(a). While the same number of random intensities are not enough to calculate $g^{(2)}(0)$ for six RGs, which can be seen from the last data point in Fig. \ref{6}(f). However, no matter what distribution the intensity follows, it is always possible to obtain $g^{(2)}(0)$ close enough to the theoretical value by employing large enough number of random intensities. Hence the image distortion in Figs. \ref{3}(e)-\ref{3}(f) can be eliminated by including more different runs to increase the number of random intensities.

\section{Conclusions}\label{conclusion}

In conclusion, we have proved that high visibility temporal ghost imaging with classical light is possible. The proposed superbunching pseudothermal light \cite{liu-arxiv,bai-2017} is helpful to improve the quality of temporal ghost image. The visibility of temporal ghost image with superbunching pseudothermal light can approach 100\% when more rotating ground glasses are employed in superbunching pseudothermal light source. The well-accepted conclusion that the visibility is different between ghost imaging with classical light and entangled photon pairs does not hold any more. The results are helpful to understand the physics of ghost imaging and increase the image quality in future temporal ghost imaging applications.

\section*{Funding}
National Natural Science Foundation of China (NSFC) (Grant No.11404255); National Basic Research Program of China (973 Program) (Grant No.2015CB654602); 111 Project of China (Grant No.B14040); Fundamental Research Funds for the Central Universities.

\end{document}